\documentclass[journal]{IEEEtran}
\usepackage[numbers,sort&compress]{natbib}
\usepackage{graphicx}
\usepackage{graphbox}
\usepackage{amsmath, amssymb, physics, bm}
\usepackage{xcolor}
\allowdisplaybreaks[4]
\ifCLASSOPTIONcompsoc
 \usepackage[caption=false,font=normalsize,labelfont=sf,textfont=sf]{subfig}
\else
 \usepackage[caption=false,font=footnotesize,subrefformat=parens]{subfig}
\fi
\usepackage{booktabs}
\makeatletter
\def\vdots@i#1#2#3{\vbox{
  #1\baselineskip#2\p@ \lineskiplimit\z@
  \kern#3\p@\hbox{.}\hbox{.}\hbox{.}}}
\DeclareRobustCommand\vdots{
  \mathchoice
  {\vdots@i{}{4}{6}}
  {\vdots@i{}{4}{6}}
  {\vdots@i{\scriptsize}{2}{1}}
  {\vdots@i{\tiny}{2}{1}}
}
\DeclareRobustCommand
  \ddots{\mathinner{\mkern0mu\raise4\p@
    \vbox{\scriptsize\kern4\p@\hbox{.}}\mkern1mu
    \raise2\p@\hbox{\scriptsize.}\mkern1mu\raise0\p@\hbox{\scriptsize.}\mkern0mu}}
\def\subtocounter#1#2{\global\advance\csname c@#1\endcsname -#2\relax}
\makeatother
\newcommand{\HH}{\mathsf{H}}

\newcommand{\jj}{\mathrm{j}}
\renewcommand{\bf}{\mathbf}

\renewcommand{\Re}[1]{\real\{#1\}}
\renewcommand{\Im}[1]{\imaginary\{#1\}}
\newcommand{\CN}{\mathcal{N}}
\newcommand{\EE}{\mathbb{E}}
\newcommand{\QQ}{\mathbb{Q}}
\newcommand{\NN}{\mathcal{N}}
\newcommand{\LU}{\mathbb{LU}}
\newcommand{\svec}[1]{\mathbf{#1}}
\newcommand{\ADC}{}
\newcommand{\AWGN}{0}

\DeclareMathOperator{\myvec}{vec}

\DeclareMathOperator{\Diag}{Diag}

\DeclareMathOperator{\Var}{Var}

\usepackage{hyperref}
\usepackage{fancyhdr}
\makeatletter
\def\ps@IEEEtitlepagestyle{%
  \def\@oddfoot{\footnotesize\mbox{}\hfill%
    \parbox{\textwidth}{%
      \centering
      © 2024 IEEE.
      Digital Object Identifier: 10.1109/LSP.2024.3356915.
      Personal use is permitted, but republication/redistribution requires IEEE permission.
      \\
      See https://www.ieee.org/publications/rights/rights-policies.html for more information.
    }%
  \hfill\mbox{}}%
}
\makeatother
\title{Symbol Detection for Coarsely Quantized OTFS}
\author{
  Junwei~He,
    Haochuan~Zhang*,
    Chao~Dong, 
    and
    Huimin~Zhu
\thanks{
J.~He and H.~Zhang are with 
Guangdong University of Technology, Guangzhou, China (e-mails: \url{sikouhjw@gmail.com}; \url{haochuan.zhang@gdut.edu.cn}).
C.~Dong is with 
Beijing University of Posts and Telecommunications, Beijing, China (e-mail: \url{dongchao@bupt.edu.cn}).
H.~Zhu is with 
Guangzhou University of Chinese Medicine, Guangzhou, China (e-mail: \url{hm_zhu@gzucm.edu.cn}).
This work was supported by Guangdong Basic and Applied Basic Research Foundation under Grant 2022A1515010196
and
Ministry of Education University-Industry Collaborative Education Program 230802345073954.
*Corresponding author: H. Zhang.%
}
}

\begin{document}

\maketitle

\begin{abstract}
This paper explicitly models a coarse and noisy quantization in a communication system empowered by orthogonal time frequency space (OTFS) for cost and power efficiency.
We first point out, with coarse quantization, the effective channel is imbalanced and thus no longer able to circularly shift the transmitted symbols along the delay-Doppler domain. Meanwhile, the effective channel is non-isotropic, which imposes a significant loss to symbol detection algorithms like the original approximate message passing. Although the algorithm of generalized expectation consistent for signal recovery (GEC-SR) can mitigate this loss, the complexity in computation is prohibitively high, mainly due to an dramatic increase in the matrix size of OTFS. 
In this context, we propose a low-complexity algorithm that embed into GEC-SR a quick inversion of the quasi-banded matrices, thus reducing the algorithm's complexity from  cubic order to linear order, while keeping the performance at almost the same level.
\end{abstract}


\IEEEpeerreviewmaketitle

\section{Introduction}
Interest in estimating signal parameters from quantized data has been increased significantly in recent years \cite{stoica2021cramer}. Ultra-wideband applications, such as millimeter-wave communications, require high sampling rates, but conventional analog-to-digital converters (ADCs) are expensive and power-hungry. In cases that are cost and power constrained, the use of high-precision ADCs is not feasible, which makes ADCs with coarse quantization a better choice for systems like 6G \cite{you2021towards}. 
For 6G a prominent waveform candidate is \emph{orthogonal time frequency space (OTFS)} \cite{hadani2017orthogonal,das2022orthogonal}, a 2D modulation technique that transforms the information to the delay-Doppler (DD) coordinate. 
OTFS enjoys an excellent robustness in high-speed vehicular scenarios, while orthogonal frequency division multiplexing (OFDM) suffers from disrupted orthogonality among subcarriers due to the high Doppler shift.

Detection for symbols in the delay-Doppler domain is key to the OTFS communications.  Linear detectors like LMMSE are usually prior-ignorant, i.e., they are unaware of the prior information of  transmitted symbols, and therefore not optimal in general sense.  Non-linear detectors like sphere decoding are optimal in detection accuracy but often suffer from an unaffordable computational complexity. An effective and efficient alternative is to use message passing (MP) for the detection of OTFS symbols, which includes:
\cite{li2023hybrid} proposed a hybrid message passing detector for fractional OTFS that combines standard MP approximate MP;
\cite{li2023otfs} adopted Gaussian mixture distribution as the messages;
\cite{li2021hybrid} designed a hybrid detection algorithm that combines MP with parallel interference cancellation;
\cite{li2021cross} detected the signals in both time and DD domains iteratively using a unitary transformation;
\cite{raviteja2018interference} developed a message passing algorithm that utilized the sparsity of the effective channel;
\cite{shan2022orthogonal} applied expectation propagation (EP) to the detection and reduced significantly its complexity by exploiting the channel structure;
\cite{yuan2021iterative} proposed an unitary approximate message passing (UAMP)  algorithm, addressing the challenge of large channel paths and fractional Doppler shifts, effectively and efficiently;
\cite{wu2021vector} circumvented the matrix inversion of vector approximate message passing (VAMP) by an average approximation.
These works, however, considered only the ideal case of infinite-precision ADCs. The influence of coarse quantization is not yet accounted for.

\begin{figure*}
  \centering
  \includegraphics[width=0.9\textwidth]{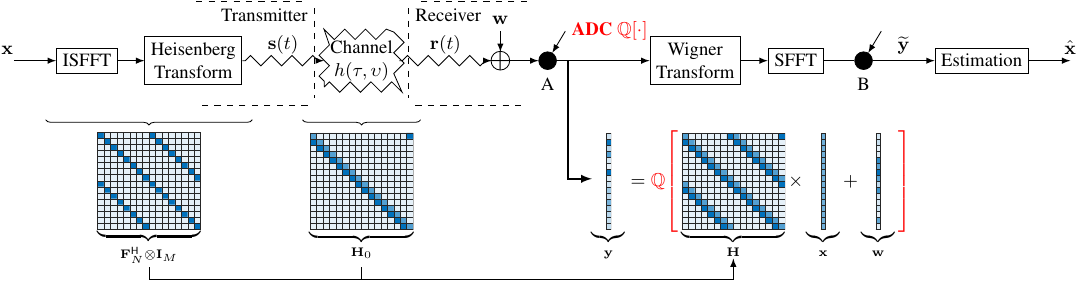}
  \caption{An example on communication via OTFS with finite-precision ADC at the receiver, where $M = 8$, $N = 2$, $P = 2$, $k_{\max} = 8$ and $l_{\max} = 2$.}\label{fig:OTFS}
\end{figure*}

This paper models explicitly the coarse and noisy quantization for the OTFS communications.
We find that a major difference between coarse quantization and the infinite-precision case is: the effective channel is no longer a multiplication of three matrices, i.e., the postprocessing transform, the multi-path fast-fading channel, and the preprocessing transform; instead, a non-linear mapping enters between the postprocessing and the remaining two, making it impossible to model them as an integrated whole. 
Ignoring the difference and applying directly the algorithms above is seen to bring about noticeable performance loss. 
To overcome the limitation, we consider a generalized linear model (GLM) \cite{rangan2011generalized} that takes in the noisy quantization, the fast-fading channel, and the preprocessing transform, and validate the performance of two efficient solvers, GAMP \cite{rangan2011generalized} and GEC-SR \cite{he2017generalized}. We find that GEC-SR is much robuster to the change of effective channel; however, the complexity of GEC-SR quickly soars up as the matrix size in OTFS squares that of the OFDM counterpart.

In this context, we propose a low-complexity GEC-SR, which utilizes a quick inversion of the quasi-banded matrices. The idea of inverting a quasi-banded matrix was not new \cite{tiwari2019low,shan2022orthogonal}; however, the channel matrix here is asymmetric (due to a lack of the postprocessing transform), and the matrix to invert is in general not quasi-banded. Interestingly, we find that if we approximate the GEC-SR covariance matrix by a scaled identity matrix, the one to invert simply reduces to be quasi-banded. It is also worth noting the method of \cite{shan2022orthogonal} is not applicable to coarse quantization, because \cite[Eq.~(40)]{shan2022orthogonal} holds only in the quantization-free case.
Finally, we carry out simulations to confirm the effectiveness of the proposed algorithm.
To sum up, we contribute in these two aspects: 
\begin{itemize}
  \item 
We point out, in the presence of coarse quantization, the effective channel becomes imbalanced, containing only one of two transform matrices, which makes the OTFS modulation unable to circularly shift the transmitted symbols along the delay-Doppler domain as designed.

  \item 
We propose a low-complexity algorithm for detecting data symbols in an OTFS system coarsely quantized, which incorporates a quick inversion of the quasi-banded matrices into GEC-SR, thus reducing the complexity from cubic order to linear order while maintaining the detection accuracy at a negligible level.
\end{itemize}

\section{
System Model for OTFS Coarsely Quantized
} \label{sec:problem-statement}
Fig.~\ref{fig:OTFS} is a block diagram of the (coarsely) quantized OTFS communication system, where $\QQ[\cdot]$ is the ADC quantization located at Position A of Fig.~\ref{fig:OTFS}. Here, the received signal is
\begin{align}
\widetilde{\svec y} 
&= 
  (\bf F_N \otimes \bf I_M) \QQ[ 
  \underbrace{\bf H_{\AWGN} (\bf F_N^{\HH} \otimes \bf I_M) }_{\bf H} \svec x + \svec w 
  ] 
  ,
  \label{eq:rx_sig}
\end{align}
where 
$\svec x = \myvec(\bf X)$ and $\widetilde{\svec y} = \myvec( \widetilde{\bf Y} )$, 
with $\bf X\in \mathbb{C}^{M \times N}$ and $\widetilde{\bf Y} \in \mathbb{C}^{M \times N}$ being the data symbols in the delay-and-Doppler domain at the transmitter and receiver, respectively, while $\myvec(\cdot)$ is the vectorization of a matrix. 
The matrix $\bf H_{\AWGN} \in \mathbb{C}^{MN \times MN}$ is the multi-path fast-fading channel.
In case of $B$-bit uniform quantization with step size $\Delta$, the mapping of $\QQ[ \cdot ]$ can be expressed as:
$
\QQ[ z] = p_1 + \Delta \sum_{i=2}^{2^B} \mathcal{H}[  z - q_i ]
$,
where 
$p_1 = \bigl( - 2^{B-1} + \frac{1}{2} \bigr) \Delta$,
and
$q_i$ is the lower limit of the $i$-th quantization interval, i.e., $q_1 = -\infty$, $q_2 = (1 - 2^{B-1}) \Delta$, $q_{i+1} = q_i + \Delta$, ($i=2,3,\cdots,2^B-1$), and $q_{2^B+1} = +\infty$.
$\mathcal{H}[z]$ is a Heaviside function, i.e., it equals $1$ if $z>0$ and $0$ otherwise.
The transitional probability density from a complex $z$ to a complex $y$ is then given by
\begin{align}\label{eq:ADC}
p( y |  z) 
&= 
  f_{\text{out}} ( \Re{y} | \Re{z})	\,
  f_{\text{out}} ( \Im{y} | \Im{z})	
\end{align}
where 
$
f_{\text{out}} (\bar{y} | \bar{z}) 
= 
\sum_i \delta(\bar{y} - p_i) \Bigl[ \Phi\Bigl(\frac{q_{i+1} - \bar{z}}{\sqrt{\sigma^2/2}}\Bigr) - \Phi\Bigl(\frac{q_{i} - \bar{z}}{\sqrt{\sigma^2/2}}\Bigr) \Bigr]
$,
$\Re{\cdot}$ and $\Im{\cdot}$ are the real and imaginary parts of a complex number, respectively,
$\delta(x)$ is the Dirac delta function, 
$p_{i+1} = p_i + \Delta$ with $i = 1,2,\cdots,2^B-1$, 
and
$\Phi(x) = \int_{-\infty}^x \NN(t | 0, 1) \dd{t}$.
The vector $\bf{w}$ is an additive white Gaussian noise (AWGN), with $\sigma^2$ being its variance.
The matrix $\bf F_M$ is the normalized $M$-point discrete Fourier transform matrix, whose $(m,n)$-th element is defined as $ \bf F_M(m,n) = \frac{1}{\sqrt{M}} e^{-\frac{2\pi \jj mn}{M}}$, 
and $\bf I_M$ is an identity matrix.
The numbers of subcarriers and time slots in the OTFS modulation are $M$ and $N$, respectively, while 
$(\cdot)^{\HH}$ is the conjugate transpose, and $\otimes$ is the Kronecker product.
$\bf{H}_0$ is modeled as \cite{yuan2021iterative,shan2022orthogonal,raviteja2018practical}:
$
\bf H_{\AWGN} = \sum_{i=1}^P h_i \Pi^{l_i} \Delta^{k_i}
$,
where 
$
\Pi = \left[\begin{smallmatrix}
  0 & \cdots & 0 & 1 \\
  1 & \ddots & 0 & 0 \\
  \vdots & \ddots & \ddots & \vdots \\
  0 & \cdots & 1 & 0
\end{smallmatrix}\right]
$ 
is an $MN \times MN$ permutation matrix, 
and 
$\Delta$ is an $MN \times MN$ diagonal matrix with non-zero elements 
$\{z^0, \cdots, z^{MN-1}\}$ with $z = e^{\frac{\jj 2 \pi}{MN}}$.
The parameter $h_i$ is the $i$-th channel gain, and  $P$ is the number of channel paths. 
This paper follows the convention of OTFS literature on symbol detection, e.g., \cite{li2021hybrid,li2021cross,yuan2021iterative}, to use a simple channel model.
More advanced models like WINNER-II and LTE-V can be considered, we leave that for further studies.

\begin{figure}[!t]
  \centering
  \subfloat[]{\label{fig:AMP+EP}%
    \includegraphics[width=.48\linewidth]{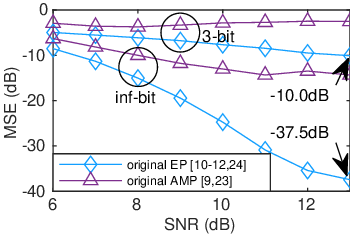}%
  }\quad
  \subfloat[]{\label{fig:GAMP+GEC}%
    \includegraphics[width=.48\linewidth]{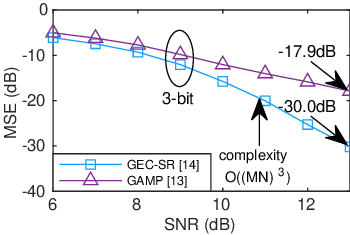}%
  }
  \caption{%
  MSE performance of state-of-the-art competitors (with $P=16$):
  (a) Original EP \cite{shan2022orthogonal,yuan2021iterative,minka2001family,wu2021vector} outperforms original AMP \cite{raviteja2018interference,donoho2009message} in the case without quantization;
  (b) Their generalized versions can handle the coarse quantization effectively, with GEC-SR \cite{he2017generalized} being the best of the four.
  }
\end{figure}

It is also worth noting the coarsely quantized model \eqref{eq:rx_sig} differs distinctly from the ideal (infinite-precision) case below 
\begin{align}
  \bf y_{\text{ideal}} 
  &=
  \underbrace{ (\bf F_N \otimes \bf I_M) \bf H_{\AWGN} (\bf F_N^{\HH} \otimes \bf I_M) }_{\bf H_{\text{ideal}}} \svec x + \svec w 
  ,
  \label{eq:ideal_model}
\end{align}
where the ideal channel matrix 
$\bf H_{\text{ideal}} 
$ is symmetric, and in effect it cyclically shifts the data symbols $\bf X$ on a delay-Doppler grid/lattice \cite{raviteja2018Low} (recall that $\bf x = \myvec(\bf X)$). 
In contrast, the introduction of $\QQ[\cdot]$ to \eqref{eq:rx_sig} breaks down the symmetry and brings in an effective channel 
$\bf H 
$ that is imbalanced.
Even in the extreme case of a single path and zero Doppler shift, i.e., $\bf H_{\AWGN} = \bf{I}_{MN}$, the two matrices are not identical: $\bf H_{\text{ideal}} \neq \bf H$; 
therefore, the quantized effective channel  $\bf H$ is no longer able to circularly shift the transmitted symbols along the delay-Doppler domain. 
To keep notations uncluttered, we perform prewhitening on $ \widetilde{\svec y}$  and obtain
\begin{align}
  \svec y 
  & = 
  (\bf F_N \otimes \bf I_M)^{-1} \widetilde{\svec y}
  =
  (\bf F_N^{\HH} \otimes \bf I_M) \widetilde{\svec y}
  ,
\end{align}
which rewrites the signal model \eqref{eq:rx_sig} in a more concise form
\begin{align}
  \svec y
  &= 
  \QQ[ \bf H \svec x + \svec w ] 
  = \QQ[ \bf z + \svec w ] 
  .
  \label{eq:new_system_model}
\end{align}
The system model \eqref{eq:new_system_model} looks similar to the quantized compressed sensing and quantized massive MIMO  \cite{wen2015joint,wen2015bayes,zou2020low,mo2014channel,liu2021decentralized}; however, the situation here is distinctly different.
Here the matrix $\bf H$ is circulant and thus correlated, but in the prior works it was assumed isotropic, e.g., i.i.d. Gaussian.
Ref. \cite{liu2021decentralized} concluded that a correlated matrix $\bf H$ will degrade the performance of the AMP-like algorithms, and this is connected to the fact that $\bf H \svec x$ is no longer i.i.d. Gaussian.
We have similar observation here: in the ideal case of Fig. \ref{fig:AMP+EP}, the original AMP  \cite{donoho2009message, raviteja2018interference} is seen to suffer from a severe performance loss, while its EP counterpart \cite{minka2001family,yuan2021iterative,wu2021vector, shan2022orthogonal} performs much better, as it treats each row of the matrix as one unit.
However, the original EP is also not as effective in the case of coarse quantization: 
changing the precision from infinite- to 3-bit will impose a  $27.5 \, \text{dB}$ loss in MSE, as evidenced by Fig. \ref{fig:AMP+EP}.
To combat coarse quantization, we apply GEC-SR algorithm \cite{he2017generalized}, an enhancement to the original EP, and find it rather robust to the change of precision, outperforming GAMP \cite{rangan2011generalized} noticeably, see Fig. \ref{fig:GAMP+GEC}.

\section{The Proposed Algorithm}\label{sec:Existing-Technics}

\begin{figure}[!t]
  \centering
  \includegraphics[width=\linewidth]{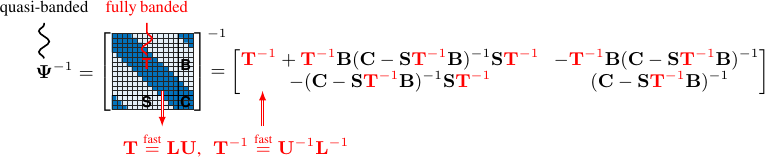}
  \caption{Lower upper decomposition on $\bf{T}$ reduces the overall complexity of inverting  $\bf{\Psi}$.
  }\label{fig:LU}
\end{figure}

Despite its robustness, the iterative algorithm GEC-SR \cite{he2017generalized} suffers from a high computational complexity, which is on the order of $\mathcal{O}(TM^3 N^3)$, where $T$ is the number of iterations.
The complexity is primarily centered on the matrix inversion
\begin{equation}
  \bigl[
    \bf H_{\ADC}^{\HH} \Diag( 1 \oslash \bf v_1^- ) \bf H_{\ADC} + \Diag( 1 \oslash \bf v_0^+ )
  \bigr]^{-1} 
  ,
  \label{eq:inverse}
\end{equation}
where $\bf v_1^-$ and $\bf v_0^+$ are parameters related to the variances, $\oslash$ is component-wise division, and $\Diag(\bf d)$ is a diagonal matrix with non-zero elements $\bf d$. 
One way to ease the burden of computation is to 
approximate the diagonal matrices by scaled identity ones and then perform an SVD on $\bf H$ before the iteration. In other words, we approximate \eqref{eq:inverse} by
\begin{equation}
  \biggl[
    \frac{ \bf H_{\ADC}^{\HH} \bf H_{\ADC} }{ v_1^- } + \frac{ \bf I }{ v_0^+ }
  \biggr]^{-1}
\!\!\! =
  (\bf F_N \otimes \bf I_M)
  \biggl[
  \underbrace{
    \frac{ \bf H_{\AWGN}^{\HH} \bf H_{\AWGN} }{ v_1^- } + \frac{ \bf I }{ v_0^+ }
  }_{
    \bf \Psi
  }
  \biggr]^{-1}
  (\bf F_N^{\HH} \otimes \bf I_M)
  \label{eq:matrix_inv}
\end{equation}
Even in this case, the complexity of $\mathcal{O}(M^3 N^3)$ \cite{fletcher2018inference} is still very high, as the OTFS matrix size has increased from $M \times N$ to $MN \times MN$, where $M \approx 50$ and $N \approx 50$, typically \cite{IEEE802.11}.
Fortunately, the matrix $\bf \Psi$ has a particular structure that admits a fast computation for its inverse.
To see this, we first note that $\bf \Psi$ is \emph{quasi-banded} \cite{shan2022orthogonal}, as shown on the l.h.s. of Fig.~\ref{fig:LU}. 
We know a \emph{fully banded} matrix has a fast matrix inversion via the lower upper decomposition (LUD), whose complexity is only $\mathcal{O}(l_{\max}^2 MN)$, with $l_{\max}$ being the maximum delay \cite{walker1989lu,shan2022orthogonal}.
However, the target matrix $\bf \Psi$ here is quasi-banded, which means applying directly on it the LUD does not help at all to reduce the complexity, because the LUD of a generic matrix is as complex as an ordinary inversion, i.e., $O(M^3 N^3)$.
Fortunately, we find that the matrix $\bf \Psi$ has its first diagonal block, denoted by $\bf T$, \emph{fully banded}.
Given the quick inverse of this fully banded block, one can readily compute the desired result utilizing the classical formula of blockwise matrix inversion. 
The cost of this last step is only $\mathcal{O}(l_{\max}^2 MN + l_{\max}^3)$ \cite{shan2022orthogonal}, where $\mathcal{O}(l_{\max}^2 MN)$ is due to matrix multiplications, and $\mathcal{O}(l_{\max}^3)$ is the ordinary inversion of $l_{\max} \times l_{\max}$ matrix $\bf{C}$ on the diagonal. 
Since $MN \gg l_{\max} $, this cost is affordable, and the overall complexity of inverting the matrix $\bf{\Psi}$ has been reduced from $\mathcal{O}(M^3 N^3)$ to $\mathcal{O}(l_{\max}^2 MN + l_{\max}^3)$.
These arguments are formalized into the following procedure:
\begin{align}
\bf{\Psi} 
&\triangleq
	\begin{bmatrix}
	\bf{T} & \bf{B} \\
	\bf{S} & \bf{C}
	\end{bmatrix}
,
\\
\bf T^{-1} 
&= 
	\bf U^{-1} \bf L^{-1}, \; \text{with} \;  
	(\bf L, \bf U) \triangleq \LU [\bf T] 
,
\\
\bf{\Delta} 
& =  
	( \bf C - \bf S \bf T^{-1} \bf B )^{-1}
,
\\
\bf \Psi^{-1}
  & =
  \begin{bmatrix}
    \bf T^{-1} + \bf T^{-1} \bf B \bf{\Delta} \bf S \bf T^{-1} & \!\!- \bf T^{-1} \bf B  \bf{\Delta} \\
    - \bf{\Delta} \bf S \bf T^{-1} & \bf{\Delta}
  \end{bmatrix} 
\\
& \triangleq
  \text{\ttfamily function\_matrix\_inverse} [\bf \Psi]
\end{align}
where $\LU [\cdot]$ is a fast LUD \cite{walker1989lu} on the (fully) banded matrix, and the subsequent matrix inversion $\bf L^{-1}$ and $\bf U^{-1}$ can also be carried out efficiently via \cite{shan2022orthogonal}.

So far, we have  reduced the complexity of the most demanding operation of GEC-SR \cite{he2017generalized} from $\mathcal{O}(M^3 N^3)$ to $\mathcal{O}(l_{\max}^2 MN + l_{\max}^3)$. 
Further replacing Eq.~(13a) and (17a) of \cite[Algorithm.~1]{he2017generalized} with  $\text{\ttfamily function\_matrix\_inverse} [\bf \Psi]$ above, we finally obtain a new algorithm, whose computational efficiency is significantly higher.
For the readers' convenience, we present this as Algorithm~1, where 
$\odot$ and $\oslash$ are component-wise product and division, respectively, and 
\begin{align*}
\EE[ \bf x | \bf m, \bf v ] 
& = 
 \frac{\int \bf x \, p(\bf x) \CN(\bf x | \bf m, \bf v) \dd{\bf x}
 }{
   \int p(\bf x) \CN(\bf x | \bf m, \bf v) \dd{\bf x}
 }
\\
\Var[ \bf x | \bf m, \bf v ] 
& = \frac{
  \int | \bf x - \EE[ \bf x | \bf m, \bf v ] |^2 p(\bf x) \CN(\bf x | \bf m, \bf v) \dd{\bf x}
}{
  \int p(\bf x) \CN(\bf x | \bf m, \bf v) \dd{\bf x}
}
\end{align*}
with $\NN (\cdot |  \bf m,  \bf v)$ being Gaussian PDF of mean $\bf m$ and covariance $\Diag (\bf v)$, 
while 
$\EE[ \bf z | \bf m, \bf v ] $ and $\Var[ \bf z | \bf m, \bf v ] $ are similarly defined, except $p(\bf x)$ is replaced by $p(\bf y | \bf z)$ and $\bf x$ by $\bf z$.

\begin{figure}
  \centering
  \includegraphics[width=.87\linewidth]{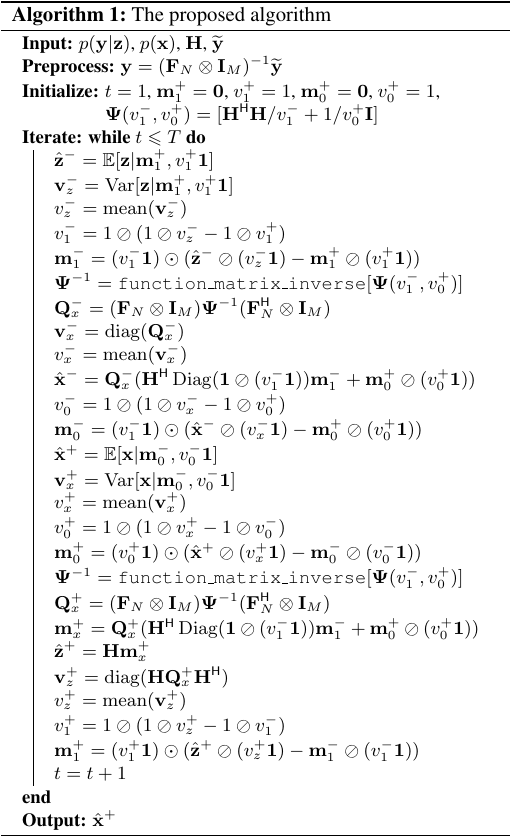}
\end{figure}

Algorithm 1 has an overall complexity of $\mathcal{O}\bigl( T \bigl( MN(l_{\max}^2 + |\mathcal{A}|) + l_{\max}^3 \bigr) \bigr)$, where $l_{\max}$ is the number of delay taps, $T$ is the number of algorithm iterations, and $|\mathcal{A}|$ is the constellation size of a modulated symbol, e.g., $|\mathcal{A}|=4$ for QPSK.
Comparing with  GEC-SR's complexity $\mathcal{O}(TM^3 N^3)$ \cite{he2017generalized}, our algorithm is far more efficient, because $\bigl( MN(l_{\max}^2 + |\mathcal{A}|) + l_{\max}^3 \bigr) \ll M^3 N^3$, given that $M$ and $N$ are both large. 
Our complexity is also much lower than other matrix-inversion-based algorithms of complexity $\mathcal{O}(M^3 N^3)$, such as classical MMSE \cite{kay1993fundamentals} and  ML-VAMP \cite{fletcher2018inference}, since, empirically, $T \approx 10$,   $l_{\max} \approx 10$, and $ T \bigl( MN(l_{\max}^2 + |\mathcal{A}|) + l_{\max}^3 \bigr) \ll M^3 N^3$.
Summing up, the proposed algorithm has the lowest complexity, as we summarize in Table~\ref{tab:complexity}.
It is also worthy of noting although Eq.~\eqref{eq:matrix_inv} here looks very similar to \cite[Eq.~(40)]{shan2022orthogonal}, the assumptions underlying these equations are very different:
\cite{shan2022orthogonal} relied on the assumption of an ideal quantization to derive a time-domain effective model \cite[Eq. (39)]{shan2022orthogonal}; in case of coarse quantization, \cite[Eq. (39)]{shan2022orthogonal} no longer holds, and the method there did not apply.
By contrast, our method applies to a much broader scope. To be specific, 
the transitional probability $p(\bf y | \bf z)$ here is not limited to coarse quantization Eq.~\eqref{eq:ADC}.
It also applies to phase retrieval \cite{schniter2014compressive}, MIMO detection \cite{yang2019symbol}, and logistic regression \cite{bishop2006pattern}.
However, due to limited space, we only present the result for coarse quantization.

\begin{table}[!t]
  \centering
  \caption{Comparison of complexity}\label{tab:complexity}
  \begin{tabular}{@{}c@{\hspace*{2.2pt}}c@{\hspace*{2.2pt}}c@{}c@{}}
    \toprule
    Algorithm   & Complexity  & Algorithm   & Complexity \\
    \midrule
    ML-VAMP\cite{fletcher2018inference}       & $\mathcal{O}(M^3 N^3)$  &
    GEC-SR\cite{he2017generalized}     & $\mathcal{O}(TM^3 N^3)$ \\
    LMMSE\cite{kay1993fundamentals}       & $\mathcal{O}(M^3 N^3)$ &
    Proposed    & $\mathcal{O}\bigl( T \bigl( MN(l_{\max}^2 + |\mathcal{A}|) + l_{\max}^3 \bigr) \bigr)$ \\
    \bottomrule
  \end{tabular}
\end{table}

\section{Simulation Results}\label{sec:simulation}

To validate the effectiveness of our algorithm, we consider the following setup:
sub-carrier number $M = 32$, slot number $N = 8$,  QPSK modulation, 
maximum delay taps $l_{\max} = 14$,
maximum Doppler shift taps $k_{\max} = 6$.
The delay index $l_i$ is randomly drawn from the integer uniform distribution $U_i[1, l_{\max}]$, with the first tap fixed at $l_1 = 0$.
The Doppler index $k_i$ is also uniformly drawn but from $U_i[-k_{\max}, k_{\max}]$, while the channel gain $h_i$ is Gaussian distributed as $\CN(0, 1/P)$.

\begin{figure}[!t]
  \centering
  \subfloat[]{%
    \includegraphics[width=.48\linewidth]{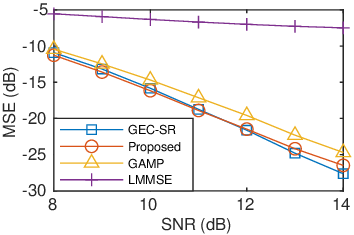}%
  }\quad%
  \subfloat[]{%
    \includegraphics[width=.48\linewidth]{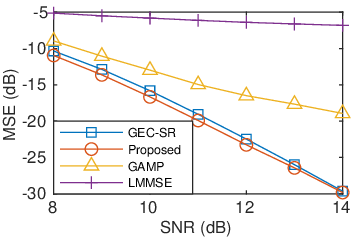}%
  }%
  \caption{The proposed algorithm is robust to the change of propagation path:  (a) $P = 6$, (b) $P = 14$.}\label{fig:ALL-SNR-MSE-3-bit}
\end{figure}

\begin{figure}[!t]
  \centering
  \subfloat[]{\label{fig:ALL-iter-MSE-3-bit}%
    \includegraphics[width=.48\linewidth]{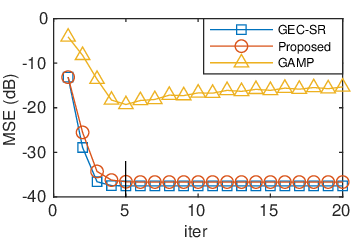}%
  }\quad
  \subfloat[]{\label{fig:mean-bit}%
    \includegraphics[width=.48\linewidth]{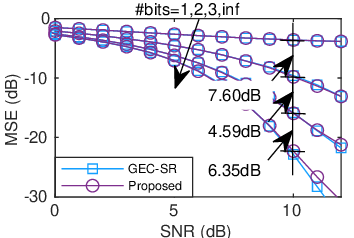}%
  }
  \caption{
    The proposed algorithm performs equally good as the $\mathcal{O}(M^3 N^3)$-complexity GEC-SR at only the cost of $\mathcal{O}(MN)$ multiplications per iteration:  	
    (a) $P = 14$, $\infty$-bit, and $\text{SNR} = 12 \, \text{dB}$; 
    (b) $P = 6$.
  }
\end{figure}

Fig.~\ref{fig:ALL-SNR-MSE-3-bit} compares the MSE performance of LMMSE, GAMP, GEC-SR and our algorithm over the entire SNR range by fixing the quantization at 3-bit. Clearly, we see that our algorithm outperforms GAMP \cite{rangan2011generalized} in both cases as the number of propagation paths $P$ increases from $6$ to $14$. Meanwhile, its performance is almost as good as GEC-SR \cite{he2017generalized}, although our complexity is significantly lower (as discussed before).

Fig.~\ref{fig:ALL-iter-MSE-3-bit} further studies the per-iteration behavior of the three, GAMP, GEC-SR, and the proposed, at 12 dB SNR. It is shown that our proposal converges very quickly: it hits the error floor within 5 iterations, which is as fast as GEC-SR. By contrast, GAMP fluctuates even after it hits the lowest point.

Fig.~\ref{fig:mean-bit} showcases the impact of quantization precision on symbol detection. We decrease the number of quantization bits from infinite to 3, 2, and 1, and see that, interestingly, empowered by our algorithm, the loss of using a 3-bit coarse quantization is only 6.35 dB, as compared to the ideal case with infinite precision. It is worthy mentioning today's communication systems typically use a 8-bit ADC \cite{stoica2021cramer}, which means there is still room for improvement in cost saving.

\section{Conclusion}\label{sec:conclusion}
OTFS is a an enabler for future wireless communications that convey information in the delay-Doppler domain.
For OTFS, this paper explicitly modeled a coarse and noisy quantization in the system.
Our contributions are two-folded:
firstly, we found that with coarse quantization, the effective channel was imbalanced and non-isotropic, which made it fail to circularly shift the transmitted symbols along the delay-Doppler domain, and also imposed a significant performance loss to the detection of symbols;
secondly, we proposed a low-complexity detection algorithm that incorporates into GEC-SR a quick inversion for the quasi-banded matrices. Our proposal can reduce the complexity from a cubic order to a linear order, while keeping the performance at the same level.

\newpage
\bibliographystyle{IEEEtran}

\bibliography{IEEEabrv.bib, ref.bib}

\end{document}